\documentstyle[aps,epsf]{revtex}
\tighten
\begin{document}
\draft

\title{Crossover Between Universality Classes in the Statistics
of 
Rare
 Events in Disordered Conductors}
\author{V. M. Apalkov and M. E. Raikh}
\address{Department of Physics, University of Utah, Salt Lake City,
UT 84112, USA}
\author{B. Shapiro}
\address{Department of Physics, Technion-Israel Institute of
Technology, Haifa 32000, Israel}

\maketitle
\begin{abstract}
The crossover  from orthogonal to the unitary
universality classes in  the distribution
of the anomalously localized states (ALS) in two-dimensional
disordered conductors
is traced as a function of magnetic field. We demonstrate that
the microscopic origin of the crossover is the change in the
{\em symmetry} of the underlying disorder configurations, that are
responsible for ALS. These disorder configurations
are of {\em  weak} magnitude (compared to the Fermi energy)
and of {\em small} size (compared to the mean free path).
We find their shape explicitly by means of the direct optimal
fluctuation method.
\end{abstract}

Statistical properties of the wave functions, $\psi (\bbox{r} )$,
 in disordered conductors
have been the subject of an intensive theoretical study
during the last decade\cite{Muzykantskii95,falko95,mirlin96}.
Considerable 
attention was devoted to the so-called anomalously localized
states (ALS). These states  constitute the  large-$| \psi |^2$
``tail'' of the wave function distribution, where  this distribution
deviates strongly from the prediction
of  the random-matrix theory. Analytical results obtained to date
are summarized in the comprehensive review
 \cite{mirlin00}. The bulk of these results are
obtained using different versions\cite{efetov,Muzykantskii'95} of
the nonlinear $\sigma $-model.
The latest works on the subject report on numerical tests of
the theoretical predictions\cite{muller97,romer98,romer00,nikolic01}.
Calculations of the ALS density, based on non-linear $\sigma$ model
\cite{Muzykantskii95,falko95,mirlin96}, yield exponentially
small tails with different numerical factors in the exponent,
depending on the universality class.
The drawback of the non-linear $\sigma$-model approach is that 
it fails to pinpoint the actual disorder realizations responsible 
for ALS. This is because within the machinery of the non-linear
$\sigma$-model the  averaging over the disorder
is performed at an early stage of the
 calculation.
In an attempt to  overcome this
deficiency, an alternative approach, 
the direct optimal fluctuation method,
was proposed in  Ref.~\onlinecite{Smolyarenko}.
In two dimensions the optimal disorder configuration obtained
in Ref.~\onlinecite{Smolyarenko} 
had a shape of a compact 
potential well with a radius of the order of the Fermi wavelength,
supplemented by a periodic modulation at large distances.
An important accomplishment of Ref. ~\onlinecite{Smolyarenko}
is the observation that, within a numerical factor in the 
exponent, the non-linear-$\sigma$-models-based results can be
reproduced by considering a simplest configuration yielding the ALS,
namely a circular (in two dimensions) region of space 
surrounded by a high and thick potential barrier.  
Then the anomalously large values of the wave function, $\psi$,
can be characterized by a single, intuitively transparent, parameter
\begin{equation}
\label{cal}
{\cal T} =|\psi _{out}|^2/|\psi _{in}|^2,
\end{equation}
where {\em "in"} and {\em "out"} refer to the inner and outer boundaries
of the barrier. 
As a direct consequence of the  compactness of the fluctuations,
the probablity of their formation is insensitive to weak magnetic
fileds, in apparent contradiction to  the  
$\sigma$-model predictions.
On the basis  of this contradiction
the authors of Ref.~\onlinecite{Smolyarenko} questioned the
ability of the non-linear $\sigma$-models to handle the ALS.

The goal of the present paper is to demonstrate that the direct optimal
fluctuation approach captures the dependence of the 
density of ALS on the universality class. Moreover, it allows to
trace the crossover between the orthogonal and unitary classes
with increasing magnetic field. We restrict our consideration to
the two-dimensional case.

The sensitivity to a weak magnetic field within the direct optimal
fluctuation approach is restored in two steps. First, instead of
circle-shape fluctuations of Ref.~\onlinecite{Smolyarenko} we consider
fluctuations of ring-shape, schematically depicted in Fig.~1. This
step is in accord with the paper by Karpov \cite{Karpov93},
who has demonstrated that in three dimensions fluctuations of a
toroidal shape, providing
a given value of $ {\cal T }$, are more probable than
spherical fluctuations, considered in Ref.~\onlinecite{Smolyarenko}.
Ring-shape fluctuations, having much bigger
area, are sensitive to the magnetic field. However, this class of
fluctuations {\em alone} does not exhibit a crossover between the
orthogonal and unitary classes. To reveal this crossover, as a
second step, we note that, on a ring, the states with angular momenta
$m$ and $-m$ are {\em degenerate} in the absence of magnetic
flux. This suggests that the fluctuations lifting this degeneracy
cause the {\em increase} of ${\cal P} (E_F, {\cal T})$, which is the
probability to find a state with energy $E_F$ and a given value of
${\cal T}$. The underlying reason for this increase is the
following. A weak coupling, $\kappa $, between the states $\left| m
\right\rangle $ and $\left| -m \right\rangle $ leads to decrease in the
formation probability of the fluctuation. This decrease is 
 {\em quadratic} in $\kappa $.
 On the other hand, the $| m\rangle$, $|- m\rangle$ splitting due to
this coupling is proportional to $|\kappa |$.
This splitting results in a
``gain'' in ${\cal T}$, which is also proportional to $|\kappa|$.
Thus, for small coupling, the ``gain'' overweighs the ``penalty'' for
creating the coupling.
 The above argument is
completely analogous to the Jahn-Teller effect \cite{Jahn}. Kusmartsev
and Rashba \cite{Kusmartsev} have employed this argument to
demonstrate that the fluctuations responsible for the {\em tail} states
of a four-fold degenerate semiconductor valence band are of ellipsoidal
rather than spherical shape. Summarizing, in a zero magnetic field
${\cal P} (E_F, {\cal T})$ is determined by the {\em warped} ring-shape
fluctuations of the random potential. In a finite magnetic field the
degeneracy $\left| m \right\rangle $, $\left| -m \right\rangle $ is
lifted; the warp does not lead to increase of ${\cal P}$
anymore. Then the fluctuations contributing to  ${\cal P}$ are close
to rings. Thus, in the statistics of rare events, the crossover between
the orthogonal and unitary classes can be viewed as switching
from {\em warped} to {\em almost perfect} rings. This is illustrated in Fig.~1(a).

The above discussion suggests the following analytical steps. Firstly,
 the fluctuation $V(\bbox{r})$ of the white-noise random potential is
 presented as a sum of the angular harmonics $V(\bbox{r}) = \sum _m
 V_{m} (\rho ) e^{im\phi} $, where $\rho = |\bbox{r}|$.  Secondly, in
 the probability ${\cal P}\{ V(\bbox{r})\}$ of the fluctuation
 $V(\bbox{r})$ determined by
\begin{equation}
\left| \ln {\cal P} \right| = \frac{\pi \nu \tau }{2\hbar } 
\! \! \int \! \! \!
d \bbox{r} V^2(\bbox{r}) =
 \frac{\pi ^2 \nu \tau}{\hbar }  
\! \! \! \!\! \sum _{m= -\infty } ^{\infty } \!  \int \! \!
 d\rho \: \rho  \left| V_{m} (\rho ) \right|^2 , \label{eq2}
\end{equation}
 only the angular harmonics $V_0 (\rho )$
 and $V_{\pm 2m} (\rho )$ are retained. This is because the
$|m \rangle $, $|- m \rangle $ coupling is provided
by the harmonics $\pm 2m$ of the random potential. In Eq.~(\ref{eq2})
$\tau $ and $\nu $ stand for the scattering time and the density of states,
respectively.
Then the system of equations relating the components $\chi _m $ and
 $\chi _{-m} $ of the wave function $\rho ^{1/2} \psi(\rho , \phi )$
takes the form
\begin{equation}
\hat{H}_0 \chi = -\frac{\hbar ^2}{2M} \frac{d^2 \chi }{ d\rho ^2 } +
     \left( V_{eff} ^{(m)}(\rho ) +\hat{V}_{2} \hat{\sigma } _{x} \right) \chi =
  E_F \chi         \label{eq3} ,
\end{equation}
where $M$ is the electron mass,
$\chi  = \left(  \chi _m ~,~ \chi_{-m} \right) $ is
 a two-component wave function, the  matrix $\hat{V}_{2} = diag (V_{2m},V_{-2m})$ is
diagonal, and $\hat{\sigma } _{i} $ are the Pauli matrices.
The effective potential in (\ref{eq3}) is a sum of a centrifugal potential and
  $V_0 (\rho )$
\begin{equation}
V_{eff}^{(m)} (\rho )  =  \frac{\hbar ^2
        (m^2 -1/4)}{2M\rho ^2}+ V_0 (\rho )  ~~. \label{Veff}
\end{equation}
The turning point, $\rho _c$, is determined by
 $V_{eff}^{(m)}(\rho _c) = E_F$ [see Fig.~1(b)].
As we will see below, the characteristic width, $w$, of the fluctuation
$V_0 (\rho )$ and the barrier thickness, $d$, [see Fig.~1(b)] are related
 as $w\ll d \ll \rho _c$.
 Similarly to Ref.~\onlinecite{Karpov93}, this separation of spatial scales
 allows a number of crucial simplifications:
(a) since $w\ll d$, it follows that, within the barrier, $V_{eff}^{(m)}(\rho )$ is
dominated by the centrifugal term; (b) since $d \ll \rho _c $, this
term can be linearized within the barrier region as 
\begin{equation}
V_{eff}^{(m)} (\rho ) = E_F + \varepsilon _0 \frac{\rho _c - \rho }{d} ~~~,
\label{eq4}
\end{equation}
where  $\varepsilon _0 \approx \hbar ^2 m^2 d /M\rho _c^3 $ is the
barrier height.
Using the above simplifications the ratio
${\cal T}= |\psi _E (\rho _c-d)|^2 / |\psi _E (\rho _c)|^2$ can be readily calculated

\begin{equation}
\left| \ln {\cal T} \right|  = 
2 \frac{\sqrt{2M}}{\hbar }
   \int _{\rho _c-d}^{\rho _c }d \rho  
 ~\left[ V_{eff}(\rho ) - E_F \right] ^{1/2}
 = \frac{2m}{3}\left( \frac{2d}{\rho _c} \right)^{3/2} =
  \frac{2m}{3} \left( \frac{\varepsilon _0}{E_F} \right)^{3/2} \label{tt} ~~.
\end{equation}

It is seen from Eq.~(\ref{tt}) that the condition $\rho _c \gg d $ can
be rewritten as $m \gg |\ln {\cal T}|$.  Equation (\ref{tt}) relates
$\cal T$ to the energy level position $\varepsilon _0$ in the
potential well centered around $\rho _c - d -w/2$ 
[see Fig.~1(b)]. Thus the
problem of finding the distribution of $\cal T$ reduces to the
conventional problem \cite{Halperin66,Zittartz66} of finding the most
probable configuration of random potential which confines the energy
level of a depth $-\varepsilon _0$.
In our case $-\varepsilon _0$ is the eigenvalue of the {\em matrix}
Hamiltonian $\hat{H} $, defined as 
\begin{equation}
\hat{H} = - \frac{\hbar ^2}{2M} \frac{d^2}{dx^2} + V_0 (x)  +
          \hat{V}_{2} \hat{\sigma } _{x}  \label{hm} ~,
\end{equation}
where $x= \rho -\rho_c +d +w/2$.
The form of the Hamiltonian Eq.~(\ref{hm}) follows from the system
Eq.~(\ref{eq3}) and the condition $w \ll d $ which allows to
neglect the change of the centrifugal potential within the potential
well [see Fig.~1(b)].

The straightforward generalization of the approach of
Ref.~\onlinecite{Halperin66,Zittartz66}
to the matrix  Hamiltonian (\ref{hm}) implies minimizing the 
functional $F\{V_0, V_{\pm 2m} \} = |\ln {\cal P} | - \lambda 
(\chi ^+ \hat{H} \chi )$.  With an appropriate choice of the Lagrange
multiplier, $\lambda $, we obtain  the
following expressions for $V_0$ and  $V_{\pm 2m}$
in terms of two-component wave function, $\chi $,
\begin{equation}
 V_0  =  \chi ^{+} \chi  ~, ~~~ V_{\pm 2m} = \chi ^{+}
  \hat{\sigma }_{\pm } \chi  ~.
      \label{Vm}
\end{equation}
At this point it is convenient to introduce dimensionless variables as
$z=x(2M\varepsilon _0)^{1/2}/\hbar $ and $\tilde{\chi }= 
 \varepsilon _0^{-1/2} \chi $. With
new notations, the matrix Schr\"{o}dinger equation, $\hat{H} \chi =
-\varepsilon _0 \chi $, reduces to
\begin{equation}
\tilde{\chi } ^{\prime \prime} + \frac{3}{2} \left( \tilde{\chi } ^{+}
\tilde{\chi }\right) \tilde{\chi }
 - \frac{1}{2} \left( \tilde{\chi } ^{+} 
    \hat{\sigma }_z \tilde{\chi }\right)
    \hat{\sigma }_z \tilde{\chi } - \tilde{\chi } =0 ~ ,
                              \label{s1}
\end{equation}
and probability (Eq.~(\ref{eq2})) of the fluctuation, providing a given value of
$\cal T$,  takes the form
\begin{equation}
\left| \ln {\cal P}_m \right|  =    \pi^2 ~ 
\frac{\nu \tau \rho_c}{\hbar } \int dx
\left( \left| V_0 (x) \right|^2 + 2 \left| V_{2m} (x) \right|^2 \right)
  =   (3/16) \pi k_F l |\ln {\cal T}| C_m ~ ,
 \label{s2}
\end{equation}
where $k_F = (2ME_F)^{1/2}/\hbar$ is the Fermi momentum,
$l=  \hbar k_F \tau /M$ is the mean free path;  dimensionless factor
$C_m$ in Eq.~(\ref{s2}) is defined as
\begin{equation}
C_m = \frac{1}{2} \int d z ~ \left[ 3 \left(\tilde{\chi }^{+}
\tilde{\chi } \right)^2 -
   \left(\tilde{\chi }^{+} \hat{\sigma }_z \tilde{\chi } \right)^2
\right] ~. \label{Cm}
\end{equation}

Equations (\ref{s1})-(\ref{Cm}) express
analytically the Jahn-Teller
physics discussed in the introduction. Indeed, this system has a
conventional instanton
solution $\tilde{\chi }_{-m} = 0$ and
$\tilde{\chi }_m = 2^{1/2}\cosh ^{-1} z$,
corresponding to the absence of warp ($V_{\pm 2m} \equiv 0$).
This solution yields $C_m = 16/3$. On the
other hand, the Jahn-Teller-type solution
$\tilde{\chi } _m = \tilde{\chi } _{-m} = (2/3)^{1/2} \cosh ^{-1} z$
leads to a {\em smaller} value $C_m=32/9$, which corresponds to
{\em exponentially} higher probability ${\cal P}_m$. Thus,
the final result for orthogonal case reads
\begin{equation}
\ln {\cal P} _m= -(2/3)\pi g |\ln {\cal T}| ~, \label{Pm}
\end{equation}
where $g=k_F l$ is the dimensionless conductance.
Remarkably, $\ln {\cal P}_m$ {\em does not} depend on the
value of the angular momentum $m$, provided that $m\gg |\ln {\cal T}|$.
 In fact, the condition $m\gg |\ln {\cal T}|$ justifies all the assumptions made
 in deriving the result Eq.~(\ref{Pm}). To see this, we write explicitly
the potentials $V_0$ and $V_{\pm 2m}$:
\begin{equation}
V_0 = 2 V_{\pm 2m} =
  \frac{4\varepsilon _0}{3\cosh ^2 \left[x(2M\varepsilon _0)^{1/2}/\hbar \right]}
~~~.    \label{V_F}
\end{equation}
Using the relation
$\varepsilon _0 = E_F \left( 3 | \ln {\cal T}| / 2 m  \right) ^{2/3}$,
it is seen  that
under the condition $m\gg |\ln {\cal T}|$ we have $V_0 (x) \ll E_F$, {\em i.e.}
the potential well is {\em shallow}.

Other assumptions used in the calculation are less restrictive.
Indeed, the typical width of the potential well $w \sim
 \hbar /(M\varepsilon _0)^{1/2} \sim k_F^{-1}(m/|\ln {\cal T}|)^{1/3}$.
On the other hand, for barrier width $d$ we have from Eq.~(\ref{tt})
$d = (\rho _c /2)
(3 |\ln {\cal T}|/2m)^{2/3}$.
Since $\rho _c = m/k_F$, we have
$d = m^{1/3}(3|\ln {\cal T}|/2)^{2/3}/2k_F$, so that
$w/d \sim |\ln ^{-1} {\cal T} | \ll 1$.
We have also assumed that the change of the centrifugal potential 
within the
 potential well
is much smaller
 than $\varepsilon _0$. 
This change can be estimated as $\varepsilon _0 w/d \sim 
\varepsilon _0 /|\ln {\cal T}|\ll \varepsilon _0$.

We now turn to the case of a finite magnetic field, $B$.
Due to the azimuthal symmetry of the problem the system of 
equations Eq.~(\ref{eq3}) relating the radial functions
$\chi_m,\chi_{-m}$ can be easily modified to
\begin{equation}
\hat{H}_0 \chi  + \frac{\hbar ^2}{2M}
 \left(\frac{m}{ l_B^2}\hat{\sigma }_z +
 \frac{\rho ^2}{4 l_B^4} \right) \chi
= E_F  \chi ~, \label{mf1}
\end{equation}
where $ l_B = (c\hbar /e B)^{1/2}$ is the magnetic length.
As can be seen from Eq.~(\ref{mf1}) there exists a following
hierarchy of magnetic field strengths. 
When $l _B \gg m^{1/2}/k_F$, then the ``confining''
term $\rho ^2/l_B^4 $ can be neglected. The latter condition
is equivalent to $R _L \gg \rho _c$, where
$R_L =k_F l_B^2$ is the Larmour radius for an electron with 
energy $E_F$.
On the other hand, it is obvious that
the term  $\pm \hbar ^2 m /(2 M l_B^2)$, lifting the $|m\rangle ,|-m \rangle$
degeneracy, becomes important when it is of the order 
of $\varepsilon _0$.
This yields $R_L \sim \rho _c (\ln {\cal T}/m)^{-2/3}$.
Since $|\ln {\cal T} |\ll m$,
we conclude that there exists an interval of magnetic fields which
affect the potential well but do not affect the barrier.
Condition $\varepsilon _0 \sim \hbar ^2 m /( M l_B^2)$ determines the
characteristic magnetic field for orthogonal-unitary 
crossover: $B_{0}^{(m)}=
(\Phi _0/2\pi) k_F^2 (3|\ln {\cal T}|/2)^{2/3} m^{-5/3}$, 
where $ \Phi _0$ is the flux quantum.
The remaining task is to find the dependence $C_m (\delta )$, where
$\delta = B/B_{0}^{(m)}$, which describes the change of $C_m$
 between ``orthogonal'', $C_m(0) = 32/9$, and ``unitary'', $C_m =16/3$, values
with increasing magnetic field, $\delta $. At finite $\delta $
 the system Eq.~(\ref{s1}) takes the form
\begin{equation}
\tilde{\chi } ^{\prime \prime} + \frac{3}{2} \left( \tilde{\chi } ^{+}
  \tilde{\chi }\right) \tilde{\chi }
 - \frac{1}{2} \left( \tilde{\chi } ^{+} \hat{\sigma }_z \tilde{\chi }\right)
    \hat{\sigma }_z \tilde{\chi } - \tilde{\chi } = \delta (1-\hat{\sigma }_z)
  \tilde{\chi },
                            \label{m1}
\end{equation}
whereas Eqs.~(\ref{s2}),(\ref{Cm}) retain their form.
At small $\delta \ll 1$ the dependence $C_m(\delta )$ can be found
by solving Eq.~(\ref{m1}) perturbatively and substituting
$\tilde {\chi }(\delta )$ into Eq.~(\ref{Cm}). This program can be carried
out analytically, yielding
\begin{equation}
C_m(\delta ) =  C_m(0) \left( 1+  \frac{3\delta }{2} \right). \label{Cmd}
\end{equation}
Our important observation is that perturbative 
expression Eq.~(\ref{Cmd})
remains valid with high accuracy up to $\delta =1/3$, when the
``unitary'' solution $\tilde{\chi }_{-m} = 0$, 
$\tilde{\chi }_{m} =  2^{1/2} \cosh ^{-1} z$ takes over.
The crossover behaviour $C_m (\delta )$ is illustrated in Fig.~2.
 In the same
figure we show the ``orthogonal'' solution of Eq.~(\ref{m1})
 at crossover point.

Overall, the direct optimal fluctuation approach yields
in two dimensions $|\ln {\cal P }| \sim g |\ln {\cal T}|$ rather than
$|\ln {\cal P }| \sim g \ln^2 {\cal T}$ in 
Ref.~\onlinecite{Smolyarenko}.
This is because the truly optimal fluctuations are close to rings
rather than to circles. The ring area, $S_m \sim \rho_c w$, being
much bigger than the area of a circle $\sim k_F^{-2} $ in
\cite{Smolyarenko} is the origin of a high sensitivity of rings to
the magnetic field. On the other hand, due to the same relation,
$S_m \gg k_F^{-2}$, the rings are much more ``vulnerable'' to
the perturbations caused by the harmonics with small angular
momenta, for which the centrifugal barrier is low.
To demonstrate this we write down
the expression for the decay rate of quasilocal state in a ring
\begin{equation}
\mbox{Im } E  =   2 \pi \sum _\mu \left(
    \langle m | V_0 | \mu \rangle ^2 +\sum _{n\neq 0}
    \langle m | V_n | \mu \rangle ^2 \right) \delta (E- E_{\mu }) 
 = \frac{\hbar }{\tau _1 } + \frac{\hbar }{\tau _2 } ~,
   \label{tau}
\end{equation}
where $\langle m | V_n |\mu \rangle $ are the matrix elements of
the harmonics $V_n (\rho )$  [see Eq.~(\ref{eq2})];
 $| \mu \rangle $ is the state of
the continuous spectrum in the absence of random potential (the corresponding
energy is $E_ {\mu}$). For simplicity in Eq.~(\ref{tau}) we 
have neglected the warp.
Above we were looking for the fluctuations with anomalously 
large $\cal T$. Note, that within a numerical factor 
${\cal T }= \hbar /\varepsilon_0 \tau _1 $.
It is seen from Eq.~(\ref{tau}) that our consideration is
justified only if $\tau _2 \gtrsim \tau _1$. 
Within the optimal fluctuation approach 
harmonics with small $n$, contributing to  $\hbar /\tau _2$,
do not affect the principal exponent. They enter at the stage of
the prefactor calculation \cite{Brezin80}. 
A {\em typical} value of $\tau _2$ is the scattering time
$\tau $, which is much shorter than $\tau _1$. The relation
$\tau _2 \gtrsim \tau _1$ is satisfied only for sparse configurations,
in which  the harmonics $V_{n} $ are {\em suppressed} within the
ring. Therefore, our result for ${\cal P}_m $ should
be multiplied by the probability, $P$,  to find such a configuration.
To calculate this probability we consider the distribution
function of time $\tau _2$:
\begin{equation}
 P (\tau_2 ) = \left\langle
 \delta \left( \frac{1}{\tau_2 } \!  - \! \! \!
 \int d\bbox{r} d\bbox{r}^{\prime } S(\bbox{r},\bbox{r}^{\prime }) V(\bbox{r})
  V(\bbox{r}^{\prime }) \right)  \right\rangle _{\! \! V(\bbox{r})} \label{P1}
\end{equation}
where
the expression for  $S(\bbox{r},\bbox{r}^{\prime })$ follows from Eq.~(\ref{tau})
\[
S(\bbox{r},\bbox{r}^{\prime }) \! = \!  \frac{2 \pi }{\hbar} \rho_c
\chi _{m}(\rho )  \chi _{m}^{*} (\rho ^{\prime })
\sum _{\mu }\psi _{\mu } (\bbox{r})  \psi _{\mu }^{*}
(\bbox{r}^{\prime })
  ~\delta\left( E - E_{\mu }\right).
\]
Averaging over random potential $V(\bbox{r})$ in Eq.~(\ref{P1}) can be
carried out explicitly.
The result is expressed in terms of  eigenvalues, $\lambda _n$, of the integral
operator with the kernel $S(\bbox{r},\bbox{r}^{\prime })$:
$\int d\bbox{r}^{\prime } S(\bbox{r},\bbox{r}^{\prime })
\phi _n (\bbox{r}^{\prime }) = \lambda _n \phi _n (\bbox{r}) $.
With the accuracy of a numerical factor the result can be obtained
qualitatively. This is because $P(\tau _2) $ is determined
by the first $N_0$ eigenvalues, which are almost equal to each other:
$\lambda _n \approx \lambda _0$ for $n < N_0$. The value $N_0$ can
be estimated as the number of squares with a side  $k_F^{-1}$
within the area of a ring $S_m$, {\em i.e.}
$N_0 \sim \rho_c w k_F^2$. On the other hand, there is an exact sum
rule $\sum _n\lambda _n = 2\pi \nu$
for the  eigenvalues $\lambda _n$. Since $ \sum _n\lambda _n
\approx \lambda _0 N_0$, we can find $\lambda _0$. This leads to
 the following result for the distribution Eq.~(\ref{P1})
$P(\tau_2) \propto  ( \nu \tau /\lambda _0 \tau _2) ^{N_0} \propto
(\tau / \tau _2)^{N_0} $.
Then the sought product $\tilde{\cal P}_m ={\cal P}_m P (\tau _2 \sim \tau_1)$,
describing the probability to find ALS with ``preexponential'' accuracy,
can be presented as
\begin{equation}
\ln \tilde{\cal P}_m  = \ln {\cal P}_m -  c_2 N_0 \ln ({\cal T} /g) 
     =   -\frac{2}{3} \pi g |\ln {\cal T}|
 - c_2 \frac{ m^{4/3}}{\ln ^{1/3} {\cal T}} \ln ({\cal T}/g),  
        \label{calPT}
\end{equation}
where $c_2\sim 1$ is a numerical factor.
For concreteness we choose in Eq.~(\ref{calPT}) the orthogonal 
expression for $\ln {\cal P}_m$.
 The above expression for $\tilde{\cal P}_m$
allows to find the optimal value of the angular momentum, $m$. 
Indeed, the
prefactor {\em falls off} rapidly with $m$, while the main 
term {\em to the first
order} is independent on $m$. We note that the $m$-dependent correction
to the principal term in $\ln {\cal P}$, which originates from the 
corrections to the barrier potential Eq.~(\ref{eq4}), is of 
the order of $|\ln {\cal P}|d/\rho_c $, i.e., it has the form
$-c_1g (\ln {\cal T}/m)^{2/3} |\ln {\cal T}|$,  where $c_1\sim 1$ is 
another numerical coefficient. 
The sign of the correction is determined by the fact that,
due to deviation of $V_{eff}^{(m)}(\rho )$ from the linear 
form Eq.~(5), preserving the value of $\ln {\cal T}$ requires 
an increase in the binding energy, $\varepsilon _0$. Thus, the 
correction is negative, and {\em increases} with $m$. 
Probability
$\tilde{\cal P}_m$ is maximal for $m=m_{opt}=c_0 (g |\ln {\cal T}| )^{1/2}$, 
where $c_0=(c_1 /2c_2 )^{1/2}$.
The resulting expression for $\tilde{\cal P}_m$ with the improved accuracy reads
\begin{equation}
|\ln \tilde{\cal P}_m| =
 \frac{2 }{3} \pi g |\ln {\cal T}| \left(1 +\frac{9~ c_0^{4/3}}{8}
  \left[ \frac{|\ln {\cal T}| }{g}\right] ^{1/3} \right) . \label{eqT}
\end{equation}
It is seen from Eq.~(\ref{eqT}) that the above consideration is valid
within the domain $1\ll \ln {\cal T} \ll g$.
For the optimal $m$ the ring radius is
$\rho _c = c_0 k_F^{-1} (g |\ln {\cal T}|)^{1/2} \ll l$, whereas
the ring width is $w=c_0^{1/3} k_F^{-1} (g /|\ln {\cal T}|)^{1/6}$.

Note in conclusion, that the  function
${\cal P}(\cal T)$ studied in the present paper describes
the  distribution of the ALS {\em widths}. 
Therefore,  among  asymptotical  distributions of various 
characteristics  of ALS (see Ref.~\onlinecite{mirlin00}), 
${\cal P}(\cal T)$ corresponds to the distribution of 
the relaxation times. 
The random potential fluctuations that determine this 
distribution are shallow compared to $E_F$. 
Whether these fluctuations emerge within $\sigma $-model-based
calculations remains to be seen.


{\em Acknowledgments}. Two of the authors MER and BS
acknowledge the hospitality of the University of K\"{o}ln and of
the Institute for Theoretical Physics at UCSB where the parts of
this work were completed.
This research was supported in part by the National Science
Foundation under Grant No. PHY99-07949.


\begin{figure}
\centerline{
\epsfxsize=4.0in
\epsfbox{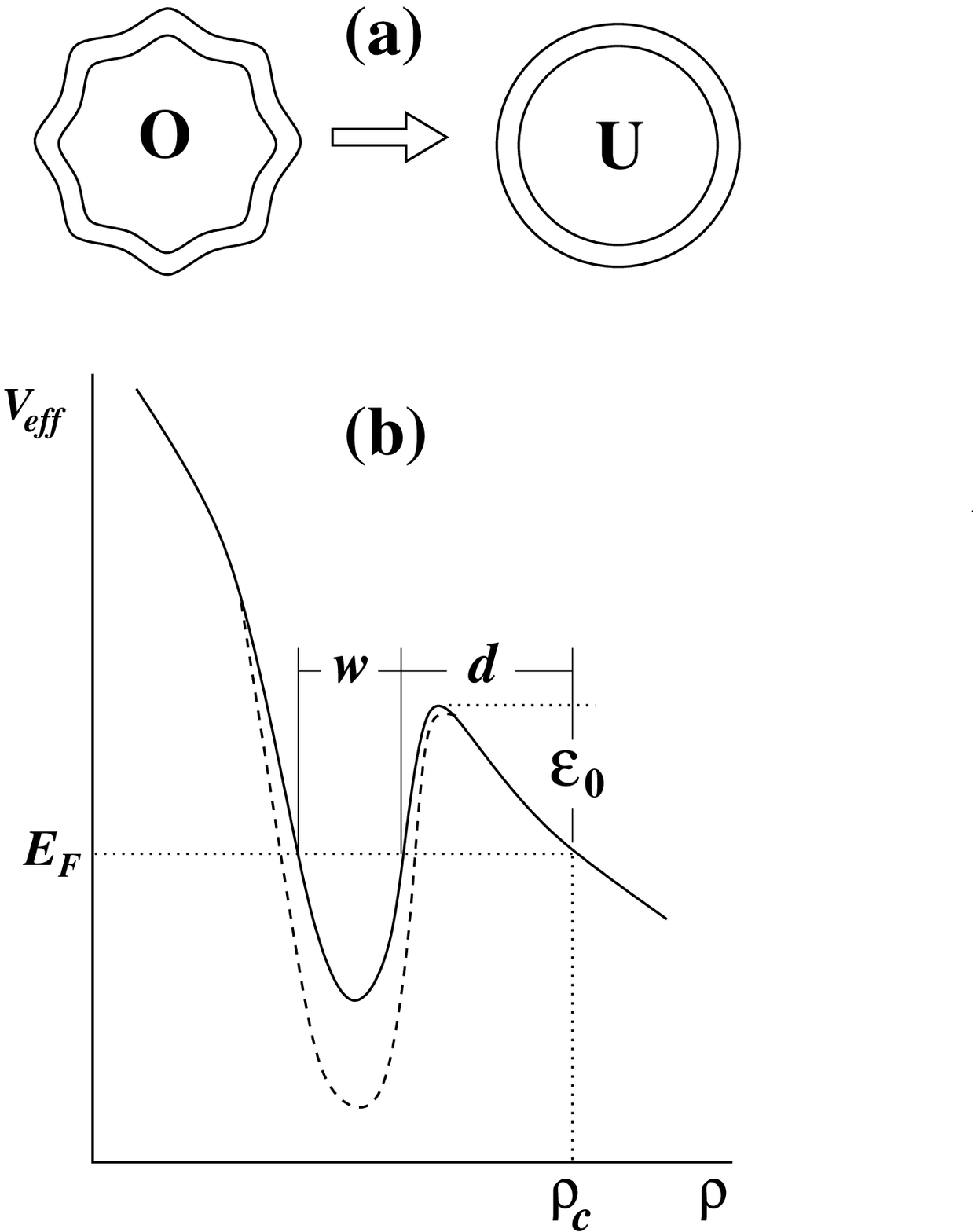}}
\vspace*{0.4in}
\protect\caption[sample]
{\sloppy{(a) Illustration of the crossover
 between orthogonal and unitary universality classes.
 Equipotential lines of the warped and ring-shaped fluctuations
 are shown schematically. (b) Effective potential
 for the angular harmonics $m$. Optimal  $V_0(\rho )$
 at  $\delta = 0$ (zero magnetic field) and at critical
 $\delta \approx 1/3$ are plotted with solid and dashed lines,
 respectively.
}}
\label{figone}
\end{figure}

\begin{figure}
\centerline{
\epsfxsize=4.0in
\epsfbox{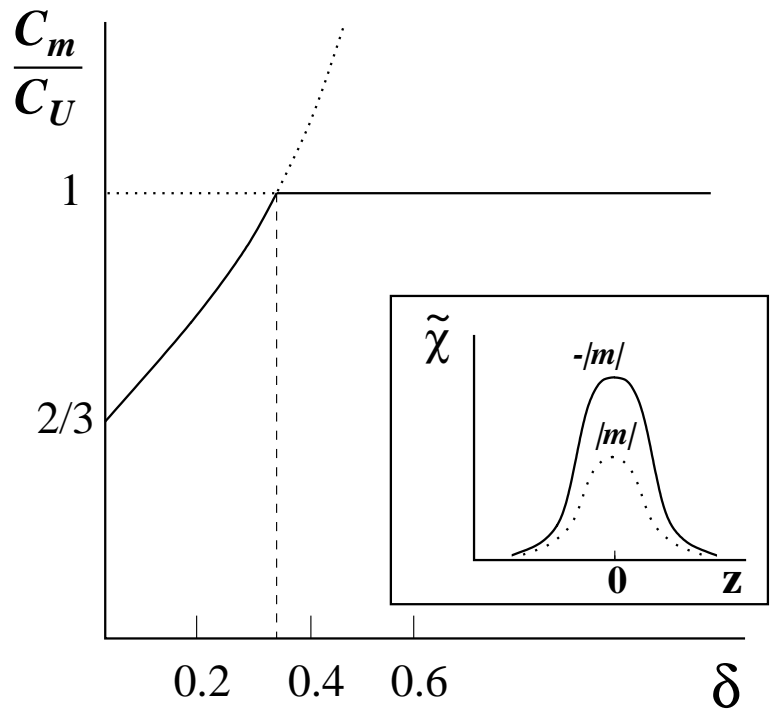}}
\vspace*{0.4in}
\protect\caption[sample]
{\sloppy{ Normalized logarithm of the  ALS density 
is shown versus the dimensionless
magnetic field, $\delta $. The inset shows the ``orthogonal'' optimal 
wave function $(\tilde{\chi } _m, \tilde{\chi }_{-m})$ 
 for crossover value $\delta \approx 1/3$. 
}}
\label{figtwo}
\end{figure}

\end{document}